\input harvmac
\Title{hep-th/9604001}
{\vbox{\centerline{$C$-theorem for two dimensional chiral theories}}}
\bigskip
\centerline{F. Bastianelli\footnote{$^*$}{email:
bastianelli@imoax1.unimo.it}}
\centerline{Dipartimento di Fisica,
Universit\`a di Modena,}
\centerline{and}
\centerline{INFN, Sezione di Bologna, Italy,}
\vglue .5cm
\centerline{and}
\vglue .5cm
\centerline{U. Lindstr\"om\footnote{$^\dagger$}{email: ul@vanosf.physto.se}}
\centerline{
Institute for Theoretical Physics, University of Stockholm, Sweden.}
\vskip 2cm

\noindent

We discuss an extension of the $C$-theorem to chiral theories.
We show that two monotonically decreasing $C$-functions
can be introduced. However, their difference is a constant
of the renormalization group flow. This constant reproduces
the 't Hooft anomaly matching conditions.

\Date{4/96}
\def\sq{\,\raise.5pt\hbox{$\mbox{.09}{.09}$}\,}

Zamolodchikov's $C$-theorem
\ref\Z{ A.B. Zamolodchikov, JETP Lett. 43 (1986) 730;
Sov. J. Nucl. Phys. 46 (1987) 1090.}
is an important result obtained in the studies of
two dimensional quantum field theories (2D QFTs). It  states that
for unitary, renormalizable, local and Poincar\'e invariant 2D QFTs
there exists a function of the couplings, $C(g^i)$, which
is decreasing along the renormalization group (RG) trajectories
leading to the infrared. Moreover, it states that
this $C$-function is stationary only at the fixed points,
where it coincides with the central charge of the corresponding
conformal field theories. The $C$-theorem makes precise the 
intuitive idea that the detailed information on the short distance 
degrees of freedom is lost under the RG flow
generated by the beta functions of the theory, and it shows that 
the RG flow is an irreversible process.

There have been several attempts directed to generalize the $C$-theorem to 
higher dimensions
\ref\C{ J.L. Cardy, Phys. Lett. B215 (1988) 749.}\ref\O {H. Osborn,
Phys. Lett. B222 (1989) 97\semi
I. Jack and H. Osborn, Nucl. Phys. B343 (1990) 647.}\ref\CFL{ A.
Cappelli, D. Friedan and J.I. Latorre, Nucl. Phys. B352
(1991) 616.}\ref\Sh{G.
Shore, Phys. Lett. B253 (1991) 380; B256 (1991) 407.}\ref\CLV{A.
Cappelli, J.I. Latorre and X. Vilas\'\i s-Cardona, hep-th/9109041,
Nucl. Phys. B376 (1992) 510.}\ref\com{J. Comellas and J.I. Latorre,
hep-th/9602123,
\lq\lq {\it Wilsonian vs. 1PI renormalization group flow 
irreversibility}\rq\rq.},
but a $C$-theorem in 4D is still missing. 
Nevertheless,
several hints that such a theorem would exist have been
obtained working in
perturbation theory \C\CLV, and checking explicit non-perturbative
examples from supersymmetric field theories 
\ref\fio{F. Bastianelli, hep-th 9511065, \lq\lq {\it Tests for C-theorems 
in 4D}\rq\rq.}.
It may help to analyze the 2D $C$-theorem from different 
perspectives, with the idea of learning something which may be useful
also in higher dimensions.

In this paper we consider explicitly the generic case of 2D QFTs which are
allowed to break parity. We will see that for chiral theories
a second $C$-function can be introduced.
However, we will show that the difference between the two $C$-functions
is constant along the trajectories of the renormalization group.
This difference measures the amount of chiral matter that is forbidden 
to become massive and decouple at low energies. Effectively,
this result reproduces the 't Hooft anomaly matching conditions 
\ref\th{G. 't Hooft, in \lq\lq{\it Recent Developments in Gauge Theories}",
Eds. G. 't Hooft et al., (Plenum Press, New York, 1980).}, here
applied to gravitational anomalies \ref\AG{L. Alvarez-Gaum\'e and 
E. Witten, Nucl. Phys B234 (1984) 269.}.
We give two different proofs of the above statements.
The first proof is obtained essentially by following 
Zamolodchikov's footsteps.
The second one is obtained by analyzing the spectral 
representation of the two-point function of the stress tensor,
as described in the nice paper of Cappelli et al.
\ref\CFL{A. Cappelli, D. Friedan and J.I. Latorre, 
Nucl. Phys. B352 (1991) 616.}.
Our interest in this second proof is to study how the extra structures 
related to chiral theories are encoded into the spectral representation.

We start with the first proof.
Let's consider a general QFT which is unitary, renormalizable, local
and Poincar\'e invariant.
Locality and Poincar\'e invariance guarantee the existence of
a conserved symmetric stress tensor $T_{\mu\nu}$:
\eqn\uno{\partial^\mu T_{\mu\nu} =0,\ \ \ T_{\mu\nu} =T_{\nu\mu}.}
We use complex coordinates\footnote{$^1$}{We work in the euclidean 
version of the theory by performing a
Wick rotation. Nevertheless, we use a language 
appropriate to the minkowskian theory.}
and denote the independent components of the stress tensor
by $T = T_{zz}$, $\bar T = T_{\bar z \bar z}$ 
and $\Theta = T_{z \bar z}$.
Then, the trace of the stress tensor is proportional to $\Theta$,
and the conservation laws read as follows
\eqn\unob{\bar \partial T + \partial \Theta =0, \ \ \ \ \  
\partial \bar T + \bar \partial \Theta =0. }
Poincar\'e invariance fixes the general form of the two
point function of the stress tensor
\eqn\due{\eqalign{ <T(z,\bar z) T(0,0)> &= 
{F(z \bar z)\over z^4},  \ \ \
<T(z,\bar z) \Theta(0,0)> = {G(z \bar z)\over {z^3 \bar z}}, \cr
<\Theta(z,\bar z) \Theta(0,0)> &= 
{H(z \bar z)\over {z^2 \bar z^2}}, \ \ \
< T(z,\bar z) \bar T(0,0)> = {E(z \bar z)\over {z^2 \bar z^2}},\cr
<\bar T(z,\bar z) \bar T(0,0)> &= 
{\bar F(z \bar z)\over \bar z^4}, \ \ \ 
<\bar T(z, \bar z) \Theta(0, 0)> = 
{\bar G(z \bar z)\over {\bar z^3  z}},\cr}}
where $F,G,H,E,\bar F,\bar G$ are undetermined 
functions of the scalar quantity $z \bar z$. 
Chiral theories are not invariant under the exchange 
$z \leftrightarrow  \bar z$. Therefore the functions $F$ and $\bar F$,
as well as $G$ and $\bar G$, are not related to each other.
Unitarity implies that the functions $F,\bar F$ and $H$ are positive, 
since they are related to the two-point function of one and the
same operator.
They can vanish only when the corresponding operator vanishes.
At the fixed point, when the correlation length become infinite,
we have a conformal field theory, 
and the trace of the stress tensor vanishes, $\Theta=0$. 
Then also the functions $G,\bar G, H$ and $E$ vanish, while 
$F={c\over 2}$ and $\bar F= {\bar c \over 2}$, 
where $c$ and $\bar c$ denote the central charges appearing
in the two copies of the Virasoro algebras generated by $T$ 
and $\bar T$, respectively.
The fact that $c$ and $\bar c$ can have a different value in the
same theory is exemplified by the 
case a free Weyl fermion which has $c = 1$ and $\bar c=0$
(or vice-versa for the opposite chirality).
This is a theory that exhibits gravitational anomalies when put
on a curved space. As a consequence, it cannot be modular invariant.
Imposing the conservation laws, eq. \unob, on the two-point functions,
eq. \due, and disregarding possible contact terms
which are not important for our considerations, we obtain
the following relations
\eqn\tre{\eqalign{ 
\dot F + \dot G &= 3 G, \ \ \ \dot G + \dot H = G + 2 H, \cr
\dot {\bar F} + \dot {\bar G} &= 3 \bar G, \ \ \ 
\dot {\bar G} + \dot H = \bar G + 2 H, \cr }}
where we have defined $\dot F = r^2 {\partial F \over \partial r^2}$
and $ r^2 = z \bar z$.
It is immediate to verify that the following combinations
\eqn\quattro{ C= 2F - 4 G -6 H, \ \ \ \bar C= 2 \bar F - 4 \bar G -6 H}
satisfy
\eqn\cinque{ \dot C = - 12 H \leq 0 , \ \ \ \dot{\bar C} =- 12 H \leq 0,}
where we have used the fact that $H$ is a non-negative quantity.
For a conformal field theory
one has $H=0$, and the two $C$-functions are stationary, 
taking the values $C=c$ and $\bar C=\bar c$.
Defining the linear combinations 
\eqn\sei{ C_\pm = {1\over 2} (C \pm \bar C),}
we see that 
\eqn\sette{ \dot C_+ = -12 H \leq 0, \ \ \ \ \dot C_- =0 .}
To complete the proof of the $C$-theorem, we first note that the
$C$-functions must have a form of the type 
\eqn\otto{C=C(r^2 \Lambda^2,g^i)}
where $\Lambda$ denotes a mass scale parametrizing the 
renormalization prescription (e.g. $\Lambda$ could be the
renormalization point, or the mass parameter
which is typically used in the minimal subtraction scheme of
dimensional regularization)\footnote{$^2$}{We 
take all coupling constants of the theory to be dimensionless
by inserting appropriate powers of $\Lambda$ .}.
In fact, the $C$-functions previously introduced must be dimensionless,
because the stress tensor is conserved and cannot develop
anomalous dimensions. In addition, it must satisfy a type
of RG equation describing the arbitrariness of $\Lambda$,
since the theory is assumed to be renormalizable.
These two properties are expressed by the equations
\eqn\nove{\eqalign{ 
&\biggl (\Lambda {\partial\over {\partial  \Lambda}} -
2 r^2 {\partial \over \partial r^2} \biggr) C= 0, \cr
&\biggl(\Lambda {\partial\over {\partial \Lambda}} + 
\beta^i {\partial \over {\partial g^i}} \biggr) C= 0, \cr }}
where $\beta^i = \beta^i (g^j)$ are the beta function of the theory,
guaranteed to exist as functions of the coupling constants
by the renormalizability assumption.
From these equations we deduce the RG equation
\eqn\dieci{ \biggl (2 r^2 {\partial\over {\partial r^2}} +
\beta^i {\partial \over {\partial g^i}} \biggr ) C= 0. }
Finally, we introduce a parameter $t$ which is increasing towards the 
infrared direction of the RG trajectory,  and integrate the 
$\beta$-function as
\eqn\undici{ {d \over {d t}} g^i = - \beta^i.}
We can then compute
\eqn\dodici{{d \over {d t}}C_+ = -\beta^i {\partial \over {\partial g^i}} C_+ 
= 2 r^2 {\partial\over {\partial r^2}} C_+ = - 24 H \leq 0,}
where we have used eqs. \undici, \dieci\ and \cinque.
Similarly
\eqn\tredici{{d \over {d t}}C_- = 0.}
Thus, we see that along the RG trajectory
there exist a monotonically decreasing 
function  $C_+$ and a constant function $C_-$.
Alternatively, we could say that there are two 
monotonically decreasing
functions, $C$ and $\bar C$, whose difference is constant along the
RG flow. $C$ and $\bar C$ reduce at the fixed point to the central charges
of the corresponding conformal field theory.
As in ref. \Z\ we could set $r=1$ in the functions 
$C,\bar C, C_+$ and $C_-$ (however this is not necessary, 
since any fixed non-zero value of $r$ would work).

Now for the second proof. We consider a spectral representation
of the two-point function of the stress-tensor, and
following ref. \CFL\ we obtain
\eqn\duno{ <T_{\mu\nu} (x) T_{\rho\sigma} (0) > = {\pi\over 3}
\int_0^\infty \! \! d\mu \ c_i(\mu) \int {d^2 p \over {(2 \pi)^2}}
\ {\rm e}^{i p x } \ {\Pi^i_{\mu\nu\rho\sigma} (p) \over {p^2 + \mu^2}}, }
where $\Pi^i_{\mu\nu\rho\sigma} (p)$ are Lorentz covariant tensors made out
of the vector $p^\mu$ and the various Lorentz invariant tensors, 
and $c_i(\mu)$ are the corresponding spectral functions.
The tensors $\Pi^i_{\mu\nu\rho\sigma}$ must respect the symmetries of 
$T_{\mu\nu}$, and reproduce the conservation equation.
The symmetries of $\Pi^i_{\mu\nu\rho\sigma}$ are:

 1. $\Pi^i_{\mu\nu\rho\sigma} = \Pi^i_{\nu\mu\rho\sigma}$,
 $\Pi^i_{\mu\nu\rho\sigma} = \Pi^i_{\mu\nu\sigma\rho}$,
which follow directly from the symmetry of $T_{\mu\nu}$.

 2. $\Pi^i_{\mu\nu\rho\sigma} = \Pi^i_{\rho\sigma\mu\nu}$, 
which can be 
deduced using causality (namely, in the spectral representation
one can use the fact that $[T_{\mu\nu} (x), T_{\rho\sigma}(0)] =0$ 
for $x$ spacelike. A short-cut is to note that causality also
allows to derive the CPT theorem, which straightforwardly implies 
the above symmetry).

Finally $\Pi^i_{\mu\nu\rho\sigma} $ must reproduce 
conservation of the stress tensor. There are two ways to achieve this
condition while respecting the above symmetries.
The first is to require that $p^\mu \Pi^i_{\mu\nu\rho\sigma}=0$.
The general solution, obtained
by using the vector $p^\mu$ and the Lorentz invariant tensors
$g_{\mu\nu}$ and $\epsilon_{\mu\nu}$, is
\eqn\ddue{\Pi^1_{\mu\nu\rho\sigma} = (p_\mu p_\nu - g_{\mu\nu}p^2)
 (p_\rho p_\sigma - g_{\rho\sigma}p^2). }
We note that after defining $\tilde p_\mu = \epsilon_{\mu\nu} p^\nu$
the above solution can be written as
\eqn\dtre{\Pi^1_{\mu\nu\rho\sigma} = \tilde p_\mu 
\tilde p_\nu \tilde p_\rho \tilde p_\sigma . }
The use of $\epsilon_{\mu\nu}$ has added no new solutions
above those given in \CFL. The corresponding
spectral function $c_1 (\mu)$, which has mass dimension 
$d=-1$, can be parametrized as
\eqn\dtrebis{
c_1(\mu) = c_1 \delta (\mu) + \tilde c_1(\mu,\Lambda), }
where $\Lambda$  is a mass scale of the theory (a term proportional to
$\mu^{-1}$ is not allowed since it would not make the  spectral
representation convergent in the infrared).
As described in ref. \CFL, the delta function term represents
the degrees of freedom at arbitrarily large distances,
while $\tilde c_1(\mu,\Lambda)$ 
is due to the density of degrees of freedom at distances $\mu^{-1}$.

A second way to achieve conservation works only at $\mu =0$. 
Namely, we require that 
$p^\mu \Pi_{\mu\nu\rho\sigma}= p^2 \Pi_{\nu\rho\sigma}$,
with an undetermined tensor $\Pi_{\nu\rho\sigma}$.
In fact, the $p^2$ factor then cancels the pole at $\mu =0$ 
in eq. \duno, and we get conservation of the stress tensor up to
contact terms (which are generically present in any case).
The independent solutions to this equation,
which do not give rise to purely contact terms,  are
\eqn\dquattro{\eqalign{
\Pi^2_{\mu\nu\rho\sigma} &= p^+_\mu p^+_\nu p^+_\rho p^+_\sigma, \cr
\Pi^3_{\mu\nu\rho\sigma} &= p^-_\mu p^-_\nu p^-_\rho p^-_\sigma, \cr}}
where $p^\pm_\mu = {1\over 2} (p_\mu \mp i \tilde p_\mu)$
are the light-cone projections of the momentum
(note that these solutions are real in the minkowskian theory).
As described, the corresponding spectral functions must be localized 
at $\mu =0$, i.e. $c_k(\mu) = c_k \delta(\mu)$ for $k=2,3$.
However, the minimum between $c_2$ and $c_3$, 
$c_{min} = { min}\{c_2,c_3\}$, does not define an universal quantity.
It can be absorbed into the constant $c_1$ of eq. \dtrebis\
by adding local terms to the effective action. This corresponds to
a redefinition of the renormalization scheme.
This statement is easily checked by
writing out explicitly the components of the various tensors
in a light-cone coordinate basis.
On the pole $p^2 = 4 p_z p_{\bar z}$ only the components
$\Pi^2_{zzzz} = p_z^4$ and
$\Pi^3_{\bar z\bar z\bar z\bar z} = p_{\bar z}^4$ 
give rise to non-local terms. When the spectral coefficients of these
two quantities are the same, one can reconstruct the tensor \ddue\
by adding local terms to the effective action.
This proves that it is consistent to set $c_{min}=0$ by
choosing a suitable renormalization scheme.
On the other hand,
the quantity $ c_2 - c_3$ has  an invariant meaning 
(i.e. independent on the renormalization scheme). 
One can immediately check that $C_- ={1\over 2}(c_2 - c_3)$.
This quantity measures the amount of chiral matter 
that is forbidden to become massive and it is
an invariant of the RG flow. In fact, the corresponding pole
present in the spectral representation \duno\ must necessarily be at 
$\mu=0$, otherwise the conservation of $T_{\mu\nu}$ is not achieved.
This massless pole is related to the gravitational 
anomalies that would arise if the theory is put on a curved background.
Thus we recognize that the RG invariance of $C_-$ is consistent with
(and required by) the 't Hooft anomaly 
matching conditions, according to which the infrared limit of
a theory must arrange itself in such a way to recreate 
the massless singularities responsible for the chiral anomalies,
the latter being typically computed in the ultraviolet regime.
Finally, a standard flowing Zamolodchikov's $C$-function
can be obtained by smearing $c_1(\mu)$ against
a density function $f(\mu)$
\eqn\dcinque{C_+ = \int d\mu f(\mu) c_1(\mu)}
satisfying the properties $f(\mu) >0$, $f(0)=1$, $f(\mu)$ 
decreasing exponentially
as $\mu \rightarrow \infty$, and $\mu {d \over d \mu} f \leq 0$.
In fact, one can compute 
\eqn\dsei{
{d\over {d t}} C_+ = - \beta^i {\partial \over {\partial g^i}} C_+  =
\Lambda {\partial \over {\partial \Lambda}} C_+ =
\int d\mu \ \tilde c_1(\mu,\Lambda) \mu {d \over {d \mu}} 
f(\mu) \leq 0,}
where we have used a Callan-Symanzik RG equation, employed dimensional 
analysis for $\tilde c_1(\mu,\Lambda)$, integrated by parts
and used unitarity that constrains the spectral functions to be
positive.
The function $f(\mu)$ can be chosen so that  $C_+$ coincides with 
the function defined in eq. \sei, (see ref. \CFL).

In summary, we have analyzed an extension of the $C$-theorem
to chiral theories. We have seen that two $C$-functions can be 
introduced, $C$ and $\bar C$, which are monotonically decreasing
along the RG flow. However, their difference is constant and it 
is related to the amount of chiral matter that cannot decouple 
at low energy by becoming massive. This result easily explains why 
't Hooft anomaly matching conditions work in this two dimensional case.
In fact, our result can be considered to be an alternative, easier
proof of such a matching.
We have also employed a spectral representation of the two point function
of the stress tensors to unearth how the extra structures
related to chiral theories are encoded into the spectral functions.

In 4D gravitational anomalies are absent as a consequence of
CPT invariance, and so it would seem that considering chiral structures
in 4D would not be of much help.
However, in our analysis we have shown how some,
apparently unrelated, things fit nicely together.
Such insight may help, after all, also in understanding
the 4D problem.

\listrefs
\end